\documentclass{jaa}
\usepackage{natbib}
\usepackage{graphicx}
\usepackage{xcolor}

\begin{document}\sloppy

\title{{\it AstroSat} Science Support Cell}


\author{Roy, J.\textsuperscript{1,*}, Alam, Md S.\textsuperscript{1}, Balamurugan, C.\textsuperscript{2}, Bhattacharya, D.\textsuperscript{1}, Bhoye, P.\textsuperscript{1}, Dewangan, G. C.\textsuperscript{1}, Hulsurkar, M.\textsuperscript{1}, Mali, N.\textsuperscript{1}, Misra, R.\textsuperscript{1}, Pore, A.\textsuperscript{1}}
\affilOne{\textsuperscript{1}Inter-University Center for Astronomy and Astrophysics (IUCAA), Ganeshkhind, Pune, 411 007, India\\}
\affilTwo{\textsuperscript{2}ISRO Telemetry, Tracking and Command Network, ( ISTRAC / ISRO ) , Bengaluru, 560058, India }


\twocolumn[{

\maketitle

\corres{jayashree@iucaa.in}

\msinfo{1 January 2015}{1 January 2015}

\begin{abstract}
{\it AstroSat} is India's first dedicated multi-wavelength space observatory launched by the Indian Space Research Organisation (ISRO) on 28 September 2015.  After launch, the {\it AstroSat} Science Support Cell (ASSC) was set up as a joint venture of ISRO and the Inter-University Centre for Astronomy and Astrophysics (IUCAA) with the primary purpose of facilitating the use of {\it AstroSat}, both for making observing proposals and for utilising archival data. The ASSC organises meetings, workshops and webinars to train users in these activities, runs a help desk to address user queries, provides utility tools and disseminates analysis software through a consolidated web portal. It also maintains the {\it AstroSat} Proposal Processing System (APPS) which is  deployed at ISSDC, a software platform central to the workflow management  of {\it AstroSat} operations. This paper illustrates the various aspects of ASSC functionality.
\end{abstract}


}]


\doinum{12.3456/s78910-011-012-3}
\artcitid{\#\#\#\#}
\volnum{000}
\year{0000}
\pgrange{1--}
\setcounter{page}{1}
\lp{1}

\section{Introduction}

{\it AstroSat} is the first Indian multi-wavelength astronomy observatory, operational for the last five years. It was launched by the Indian Space Research Organisation (ISRO) on 28 September 2015 at 10:00 AM and placed at an altitude of 650 km above the Earth into a circular orbit at an inclination of 6$^{\circ}$ and an orbital period of 98 min. {\it AstroSat} operates in specific bands over the energy range from $\sim$ 1eV to 100 keV by means of 4 co-aligned instruments, one UV imaging telescope and three co-aligned X-ray astronomy instruments. The Ultra Violet Imaging Telescope (UVIT) telescope \citep{tandon20, subramanium16} is capable of simultaneous observations in the optical, near and far UV bands. A Soft X-ray Telescope (SXT) \citep{singh17, singh16} observes in the 0.3-8 keV band. The hard X-ray instruments Large Area X-ray Proportional Counter (LAXPC) \citep{antia17, agrawal17} and Cadmium Zinc Telluride Imager (CZTI) \citep{rao17, chattopadhyay16, vadawale15} are sensitive in the 3-80 keV and 10-150 keV bands respectively.
 There is also a Scanning Sky Monitor (SSM) \citep{ramadevi17, ramadevi2017} to monitor variability of cosmic X-ray sources in 2.5-10 keV. With five specialized instruments on-board, {\it AstroSat}'s mission is to observe a wide variety of astronomical sources such as star-forming regions, violent explosions like Gamma-ray bursts and high energy emission from systems harbouring black holes, neutron stars, white dwarf.
For the last five years, {\it AstroSat} has been producing unprecedented high-quality data from cosmic sources. It is a proposal-driven, observatory class mission, with a large fraction of the observing time open to national users. {\it AstroSat} has already observed $\sim$ 1200 distinct targets in $\sim$ 2100 pointings. Moreover, 239 observations of Target of Opportunity (TOO) have been carried out. These observations are providing a unique opportunity to a large number of scientists from all over the country to carry out front-line research using a state of the art indigenous national facility. {\it AstroSat} has huge potential to attract young students and researchers in Indian Universities and institutions and engage them in high-quality astronomical research. Moreover, young teachers in Universities and Colleges would also be interested in using {\it AstroSat} observations. However, since this is the first time such an observatory is available in India, the young scientists would necessarily require initiation, training and mentoring, if they have to accurately utilize the {\it AstroSat} opportunity. There are already $\sim$ 1369 users {\it AstroSat} from 42 different countries. There is a critical need to address the queries from these users and to facilitate optimal usage of {\it AstroSat} data.

To address these requirements, the Indian Space Research Organization (ISRO) and IUCAA have established the {\it AstroSat} Science Support Cell (ASSC)\footnote{http://{\it AstroSat}-ssc.iucaa.in/}. The cell which is hosted at IUCAA started its operations in May 2016. With most of the {\it AstroSat} data made available publicly and more such releases expected in the near future, the role of ASSC is critical in providing the science support. We describe the main objectives and functionalities of ASSC in assisting Indian and International community to participate in observations and using data from {\it AstroSat}.

\section{Functionalities of ASSC}

\subsection{Overview of the web-portal}

ASSC webpage serves as a web repository of all the software, latest updates, proposal cycles, publications and link to various webpages  relevant for proposing with {\it AstroSat} and data analysis. ASSC has a centralized web portal which provides AstroSat users with links to tools developed at ISRO, Payload Operation Centres (POC) as well as documents/tools useful for scientific analysis. The Document section in Proposers page allows user to access the Proposers Guide which explains the details of the {\it AstroSat} proposal preparation and submission procedure using the APPS software, it also acts as the proposers’ manual for the APPS software. There is a Red Book available in the Document section which gives the details of all the accepted AO and GT proposals along with the abstract. The Redbook helps to check if there is already an observation proposed or completed for the source of interest that the proposer wants to propose. The home page of the web portal provides the {\it AstroSat} schedule viewer link\footnote{https://webapps.issdc.gov.in/MCAP/} to check the status of a proposed observation or the planned scheduling of the observations. Further, {\it AstroSat} handbook is provided to the users explaining the technical details about the design and characteristics of each of the payloads on-board the satellite, instrument calibration, selection of filters, different modes of operations, the primary scientific objectives intended with these payloads. The relative angle between the payload boresight’s\footnote{http://{\it AstroSat}-ssc.iucaa.in/?q=documents} (UVIT, SXT, LAXPC) before and after alignment corrections is also available which enables the proposer to account for the offset of secondary instruments for the proposed observation with the choice of primary instrument. After a successful observation of a proposed target the procedure of the data processing, data rights, proprietary period, etc. are provided in the document "{\it AstroSat} data usage guidelines". The ASSC web portal maintains up to date information and online tools to help users propose for {\it AstroSat} observations (explained in details in the next subsection on Software and other resources for proposal writing). The proposers page provides the pieces of information on the upcoming, ongoing and past announcements of opportunities cycles, scientific and technical proposal templates, online tools such as exposure time and visibility calculators, simulation and necessary documents such as the proposers' guide. For data processing and analysis, the latest available pipelines for all four instruments are maintained including sample data which allows the user to actually try out the analysis (explained in details in the next subsection of Software and other resources for data analysis). The results of the analysis of the sample data are made available allowing the new user to evaluate whether his/her analyses match with the standard ones or not. Additional important software such as the {\it AstroSat} orbit file generator, the Barycentric correction code and the {\it AstroSat} Time converter developed in-house at the ASSC are made available through the portal. ASSC has also developed advanced scientific tools such as to compute frequency and energy-dependent time lag from LAXPC data.
The "Recent Updates" panel in the right side of the page informs all the users about the recent updates and announcements relevant to {\it AstroSat} for example recent software updates, proposal cycle deadlines, health of any instrument, etc. The home page of the webpage displays all the publications based on {\it AstroSat} data from all the instruments. Home page also announces the AO cycle opening from time to time as advertised by ISRO. ASSC webpage also displays the picture of the month based on {\it AstroSat} results which is regularly updated.

\subsection{Software, webTools and resources}

\subsubsection{Software and resources for proposal preparation}

ASSC provides the software, online/offline tools, documents, etc., necessary for the proposal writing. It is a single platform which contains proposal preparation materials for all the instruments UVIT, SXT, LAXPC, and CZTI onboard {\it AstroSat}. Using these materials, users can simulate energy spectrum, power spectrum, image, etc. They can find the optimum value of exposure time to achieve a scientific goal with {\it AstroSat} observations. As the satellite time is very precious, it is a crucial step in proposal writing and acceptance of the proposal. These proposal preparation tools are described below in more details.

\begin{itemize}
\item \textit{{\it AstroSat} Visibility:} ASSC provides {\it AstroSat} visibility calculators i.e \texttt{ASTROVIEWER} developed by ISRO\footnote{https://webapps.issdc.gov.in/astroviewer/jsp/UserInput.jsp} and \texttt{AVIS}\footnote{http://{\it AstroSat}-ssc.iucaa.in:8080/AstroVisCal/} developed by ASSC, which tell the user whether a target of interest will be visible to {\it AstroSat} for observations during specific periods. There may be earth occultation when user wants to observe the source. \texttt{ASTROVIEWER} incorporates visibility constraints arising from the Sun, the Moon, the Earth and the telescope Ram angle in the orbit.  \texttt{AVIS}, in addition, also provides instrument-specific visibility.
\item \textit{Bright source checking tools:} For the safety of the UVIT detector, it is recommended to check for the presence of bright sources in the Field of View (FOV) which may damage the instrument irrespective of whether these bright sources are the subject of the proposal or not. ASSC provides link to a tool, called Bright Source Warning Tool (\texttt{BSWT})\footnote{https://uvit.iiap.res.in/Software/bswt}, developed by UVIT POC. This tool generates the list of bright visible sources with a magnitude which are near the proposed position. It also mentions that whether any source is too bright to observe or not and which filter is not safe for the observation. This tool tells about the safety of UVIT filter based on the modelling from visible magnitude. There are other tools to check the safety of VIS and FUV/NUV channels of UVIT separately. \texttt{UVIT VIS filter checking tool (Theia)}\footnote{http://uvit.iiap.res.in/Software/theia/} is used for VIS filter whereas a tool \texttt{UVIT FUV/NUV filter checking tool (Gaia)}\footnote{http://uvit.iiap.res.in/Software/gaia/} is used for FUV and NUV filters. Both are developed by UVIT POC. ASSC also provides a \texttt{UVIT 9-point coordinate generator script}, developed by ASSC itself. This tool generates 9 points around the proposed source to quickly scan the area around the proposed target. This is essential for those fields for which prior UV observations do not exist.

\item \textit{Exposure Time Calculator for UVIT instrument:} ASSC provides two separate online tools i.e , \texttt{UVIT exposure time calculator at IIA}\footnote{https://uvit.iiap.res.in/Software/etc} and \texttt{UVIT exposure time calculator at Calgary} \citep{leahy14} \footnote{http://uvit.ras.ucalgary.ca/cgi-bin/UVIT.cgi}, to estimate the exposure time. This tool calculates the exposure time by taking source characteristics as input. It gives the exposure time to achieve a particular signal to noise ratio. If a user provides the exposure time as input then, in that case, it will give the signal to noise ratio in the output.
\item \textit{WEBPIMMS:} For observations of X-ray sources, if a user has prior knowledge of some spectral characteristics of the source with other space-based telescopes like XMM-Newton, Chandra, etc., or with the same telescope {\it AstroSat}, the online tool \texttt{WEBPIMMS}\footnote{http://{\it AstroSat}-ssc.iucaa.in:8080/WebPIMMS\_ASTRO/index.jsp}, developed by ASSC based on HEASARC tools predicts the observable count rates based on the spectral properties provided by the user. This tool can be used for instruments SXT, LAXPC, CZTI and SSM. It uses some simple inbuilt models.
\end{itemize}
If a user wants to simulate spectrum of a source with any of the {\it AstroSat} instrument, ASSC provides two softwares i.e \texttt{{\it AstroSat.sl}} and \texttt{Event\_Simul}, developed by ASSC, for spectral and temporal data respectively. For UVIT, ASSC also provides a tool \texttt{UVIT$\_$simulator}, developed by UVIT POC, to simulate image. All these software tools can be found at the section "downloadable resources" of the webpage "proposal preparation"\footnote{http://{\it AstroSat}-ssc.iucaa.in/?q=proposal\_preparation}.
\begin{itemize}
\item \textit{Spectral simulation code {\it AstroSat}.sl:}  An ISIS code \texttt{{\it AstroSat}.sl}, can be used to simulate the energy spectrum using any model. It is a single code which can work for all instruments UVIT, SXT, LAXPC and CZTI. Users can get the expected count rate for input spectral parameters for all the instruments. However, a user can take response files from this place and perform the simulation on another platform like XSPEC, etc. also.
\end{itemize}
 
\begin{itemize} 
\item \textit{Timing simulation code Event\_Simul}: For the simulation of temporal properties like lightcurve, power spectrum, detection of quasi-periodic oscillations (QPOs), etc., ASSC provides a simulation code \texttt{Event\_Simul}. It works for the LAXPC instrument.

\item \textit{UVIT Image simulation code:} An image simulation code, \texttt{UVIT\_simulator}  to simulate the image of the source. In this software, a user provides the image of the same source with other telescopes and some parameters like exposure time, size of the field of view in arcsec, effective area of UVIT detector, etc. as input.

\end{itemize}

 Finally, a user can prepare a proposal using the tools mentioned above. The user needs to prepare the proposals using scientific justification and technical justification templates and submit the proposal through the {\it AstroSat} Proposal Processing System (APPS).

\subsubsection{Software and resources for data analysis} 

Since photons from an astronomical source can interact with the instrument, background photons can also reach on the detector. We should have some procedure to differentiate between source$'$s genuine feature and unwanted signals. We also need some methods to convert the observational data into a particular format, which can be used to generate spectra, images, etc. For this purpose, ASSC brings necessary software tools like pipeline software, calibration files, etc., from the respective POC of instruments. It also develops some tools. It also has some data analysis tools. The ASSC web-portal makes available these tools and related documents. All these materials can be found on the webpage "Data \& Analysis"\footnote{http://{\it AstroSat}-ssc.iucaa.in/?q=data\_and\_analysis}. ASSC team updates the materials from time to time. There are three software that can be used to analyze LAXPC data.
\begin{itemize}

\item\textit{LAXPCsoftware} FORMAT A software to do all tasks separately.
\item\texttt{laxpc\_soft} FORMAT B software to do all tasks in one step.
\item\textit{LAXPC Level 2 Data Pipeline Ver 3.1} is another package to analyse LAXPC data. 
\end{itemize} LAXPC gain record and HV record are also updated regularly.
For SXT data analysis, other than pipeline software, ASSC provides software to merge orbit wise event files and to generate the ancillary response file (ARF) file formation \footnote{http://{\it AstroSat}-ssc.iucaa.in/?q=sxtData}. These software are also developed by SXT POC.

ASSC also provides a higher-level resource for data analysis. A software \texttt{COMPT-Time-Lag-RMS}, developed by ASSC, is used to study the dependence between spectral and temporal properties. It can compute the energy-dependent time lag when seed photon temperature is varied or coronal heating rate is varied. Other than this, ASSC provides an online tool to generate an orbit file\footnote{http://{\it AstroSat}-ssc.iucaa.in:8080/orbitgen/} which is useful in the barycentre correction. Software related to the barycentre correction (as1bary) is also provided \footnote{http://{\it AstroSat}-ssc.iucaa.in/?q=data\_and\_analysis}. An online tool for the conversion between {\it AstroSat} time and other time (like Julian date, modified Julian date, etc.) and vice versa is also available here\footnote{http://{\it AstroSat}-ssc.iucaa.in:8080/{\it AstroSat}time/}.

Documents related to the installation of pipeline software and analysis procedure are either attached with the software or separately available. As for UVIT data analysis, two documents \texttt{UVIT\_Pipeline\_Cookbook\_v5.pdf} and \texttt{UL2P\_quick\_installation\_and\_output\_product\_ help\_v9.pdf}\footnote{http://{\it AstroSat}-ssc.iucaa.in/?q=uvitData} can be used to understand the procedure. At the same time, the SXT pipeline guide is attached to the pipeline software. For LAXPC data analysis software LAXPCSOFT, readme files for all the tasks are attached to the software. A user can follow these files to understand the analysis. For the CZTI, ASSC provides the user guide for the same.

\subsection{APPS}

APPS\footnote{https://apps.issdc.gov.in/apps/auth/login.jsp} is an online webpage assisting scientists in proposal preparation, submission, scientific and technical review and selection process. APPS caters to different types of users including the General or Guest Observers, Payload Operation Centre team members, payload scientists and proposal reviewers. APPS is deployed at the Indian Space Science Data Centre (ISSDC), ISRO, Bangalore. APPS has been successfully used for proposal preparation, submission and selection for observations in the (i) performance verification (PV) phase in the first six months of {\it AstroSat} operations, (ii) six Guaranteed Time (GT) cycles, (iii) ten Announcement of Opportunity (AO) Cycles, Target of Opportunity proposals, Calibration proposals, and Legacy proposals. Details of the APPS is provided in \citep{bala20} of this issue. Over the last five years, the ASSC at IUCAA has played a crucial role in fixing of bugs and security issues, documentation, and feature enhancements. They have visited ISSDC, Bangalore and coordinated with the ISRO team to make APPS software secure and on managing resources and memory leakage in APPS software.

\subsection{{\it AstroSat} Helpdesk and Visitor Programmes}

The ASSC team at IUCAA manages {\it AstroSat} helpdesk\footnote{{\it AstroSat}help@iucaa.in} to address a large number of queries from {\it AstroSat} users on proposal preparation, software installation and usage and data analysis. 
These queries are based on the issues faced by the proposers in procuring data based on the announcement of opportunity and target of opportunity observation cycles, issues related to software installation and analysis procedure, addressing software related bugs. Helpdesk re-directs the more involved inquires to the respective Payload operation
centres. {\it AstroSat} helpdesk facilitates the Indian and international proposers with proposal based queries and ensures smooth proposal submission during the proposal cycles announced at regular intervals from ISRO. Some of the queries which are generic to all the users are listed in the frequently asked questions (FAQ\footnote{http://{\it AstroSat}-ssc.iucaa.in/?q=faq}) list available at ASSC webpage. \\
In the last 4 years, the ASSC hosted more than $\sim$300 visits by Ph.D students, University and College teachers, as well as a few experts who interacted with the ASSC staff and used its facility. Help was provided to these visitors, who came from different parts of the country, to analyze the data they had from their own proposals or given to them.

\subsection{Workshops and Meetings organized by ASSC}
\begin{figure*}
\begin{center}
{\includegraphics[scale=0.5]{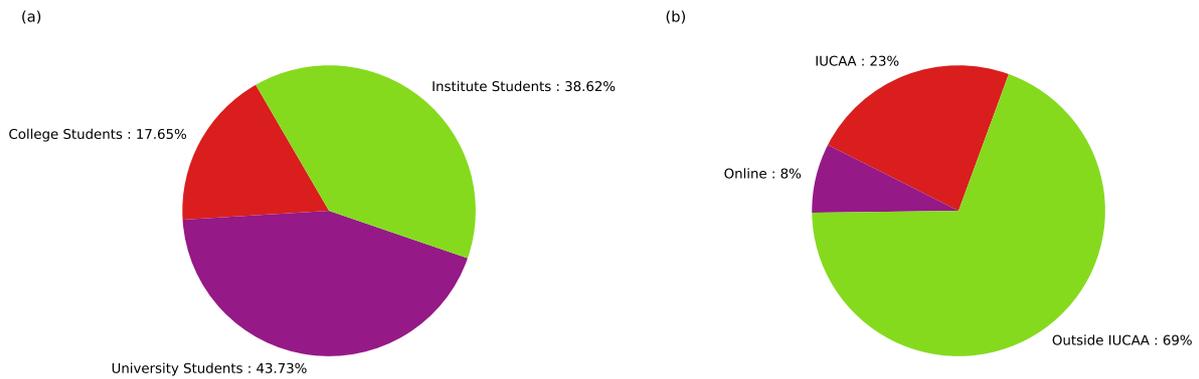}}
\end{center}
\caption{Pie chart (a) shows the distribution of participants affliations of undergraduate colleges affiliated to universities, universities and research institutions, (b) presents the locations of different workshops organized by ASSC.}
\label{workshop}
\end{figure*}

ASSC organizes national workshops and meetings in various locations across India since July 2016. These workshops were a set of extensive demonstration sessions where the participants were instructed on how to propose for {\it AstroSat} observations, in particular the choice of how to configure each instrument, exposure times, filters etc., in order to optimize the Science output. Proposal writing techniques, simulation of expected science results were explained. Other than the scientific and technical justification details and types of proposals to be sought for during an announced cycle, these workshops focus on hands-on-training sessions of data analysis techniques of different instrument onboard {\it AstroSat}. The content of the webinars and a few talks are also available in the web-portal. ASSC also conducts refresher course in Astronomy and Astrophysics for teachers at Indian universities and colleges for observational and theoretical aspects of astronomy and state-of-the-art methods of data analysis, especially with {\it AstroSat}. ASSC trained more than 489 participants to enable them to propose with {\it AstroSat} and to handle data analysis of {\it AstroSat} during its 4 years of existence. A Workshop was conducted where experts guided M.Sc/M. Phil students to analyze and interpret {\it AstroSat} data, which directly led to the publication of the results with the students as co-authors Mudambi et al. 2020 and Jithesh et al. 2019. Details of completed and upcoming workshops are regularly updated in the webpage http://astrosat-ssc.iucaa.in/?q=workshops. The distribution of participants’ affliations and location of the workshops organized by ASSC is shown in Figure \ref{workshop}. {\it AstroSat}  related meetings such as those of the time allocating committee and the Science Working group are also organized by the ASSC.

\subsection{Resources relevant to publication and calibration}

The calibration webpage of ASSC website provides the list of calibration sources observed by the {\it AstroSat} satellite. The list mentions the effective exposure of each of the observation instrument wise. It is highly beneficial for the user to use this data for calibration purposes. Some of these datasets are open access and some proprietary of the proposer or the instrument teams. This page also provides the list of publications based on calibration and performance of the payloads onboard the {\it AstroSat} satellite.\\

With a lot of the {\it AstroSat} data being released to the public and more such releases expected in the near future, the ASSC will be critical in providing the opportunity and means for the larger scientific community to take part in optimal utilization of {\it AstroSat} data.

\section*{Acknowledgements}
ASSC team would like to thank {\it AstroSat} mission of the Indian Space Research Organisation (ISRO), the Indian Space Science Data Centre (ISSDC) team. We thank all the instruments POC teams for their support for solving queries and time to time help with new software updates.

\end{document}